\pdfoutput=1
\documentclass[conference]{IEEEtran}
\IEEEoverridecommandlockouts
\usepackage[utf8]{inputenc}
\usepackage[T1]{fontenc}
\usepackage{amsmath,amssymb,bm}
\usepackage{booktabs,multirow,array}
\usepackage{graphicx}
\usepackage{tikz}
\usetikzlibrary{positioning,arrows.meta}
\usepackage{cite}
\usepackage{xurl}
\usepackage{microtype}
\usepackage{balance}
\usepackage{hyperref}

\newcommand{\stopgrad}{\operatorname{sg}}
\makeatletter
\g@addto@macro\normalsize{%
  \setlength\abovedisplayskip{4pt plus 1pt minus 2pt}%
  \setlength\belowdisplayskip{4pt plus 1pt minus 2pt}%
  \setlength\abovedisplayshortskip{2pt plus 1pt}%
  \setlength\belowdisplayshortskip{2pt plus 1pt}%
}
\makeatother
\begin{document}
\title{Fr\'echet Distance Loss on Speech Representations for Text-to-Speech Synthesis\thanks{Code and audio samples are available at \url{https://github.com/voidful/fd-speech} and \url{https://huggingface.co/spaces/voidful/fd-speech-demo}.}}
\author{%
\IEEEauthorblockN{Ho-Lam Chung$^{1}$,
Kuan-Po Huang$^{2}$,
Bo-Ru Lu$^{\dagger}$\thanks{$^\dagger$This work is unrelated to the author's position at Amazon.},
Hung-yi Lee$^{1}$}
\IEEEauthorblockA{\textit{$^{1}$Graduate Institute of Communication Engineering, National Taiwan University, Taipei, Taiwan}\\
\textit{$^{2}$Graduate Institute of Computer Science and Information Engineering, National Taiwan University, Taipei, Taiwan}\\
\textit{$^\dagger$Amazon}}
}
\maketitle
\begin{abstract}
Few-step diffusion and flow-matching text-to-speech (TTS) models are usually trained with local objectives, such as conditional flow matching, reconstruction, and stop prediction. These losses provide stable optimization, but they never ask whether sampled speech follows the distribution of high-quality speech. We propose Speech Representation Fr\'echet Distance loss (SR-FD), a training-time distributional regularizer for tokenizer-free flow-matching autoregressive TTS. During fine-tuning, the model synthesizes speech with the same few-step sampler used at deployment, and SR-FD matches the mean and covariance of frozen Whisper and CTC features of this speech to reference statistics computed offline from three complementary content targets. The loss requires no discriminator and no inference-time computation. On Seed-TTS English, four-step SR-FD fine-tuning reduces word error rate (WER) from the original four-step VoxCPM2 baseline's 2.23\% to 1.41\%, a 36.5\% relative reduction, and also surpasses the original ten-step baseline at 1.74\%; both gains are significant under an utterance-level paired bootstrap. Speaker similarity and objective quality proxies are close to the ten-step level, and a blinded listening test finds no reliable listener preference against the ten-step baseline. An error analysis attributes the gain to fewer content substitutions across all prompt lengths. We frame SR-FD as an intelligibility regularizer for few-step TTS, not a general quality objective.
\end{abstract}
\section{Introduction}
Diffusion and flow-matching TTS systems achieve high synthesis
quality~\cite{le2024voicebox,shen2024naturalspeech2,ju2024naturalspeech3,chen2024f5tts}, but they require multiple numerical integration steps at inference, imposing deployment latency. Reducing the number of flow-matching integration steps is therefore a practical strategy for real-time use. Standard training
objectives, however, operate locally: the flow-matching loss supervises a per-frame velocity field, reconstruction losses target individual acoustic frames, and the stop-prediction loss governs utterance length termination~\cite{voxcpm2repo}. Teacher-distillation
methods~\cite{salimans2022progressive,song2023consistency} reduce the
trajectory-level step-count gap, but they do not directly constrain the
population distribution of complete utterances that the compressed sampler produces. A model may therefore achieve low training loss while the distribution of its sampled outputs drifts from that of full-step inference in content, duration, prosody, and acoustic quality.
This issue is especially visible in tokenizer-free flow-matching autoregressive TTS models. VoxCPM2~\cite{voxcpm2repo} generates continuous acoustic representations rather than discrete codec tokens, avoiding quantization artifacts of discrete-based systems, but when its sampler is compressed to four steps, the generated distribution can drift from the one generated by more sampling steps. In our Seed-TTS English evaluation~\cite{anastassiou2024seedtts}, the original VoxCPM2 four-step sampling baseline has a substantially worse word error rate (WER) than the original ten-step baseline, showing that the low-step sampler is not merely a faster version of the same model.
We propose Speech Representation Fr\'echet Distance loss (SR-FD), a
training-time distributional regularizer for tokenizer-free flow-matching autoregressive TTS. During fine-tuning, SR-FD synthesizes speech with the same flow-matching sampler used at deployment, extracts frozen speech representations from the generated audio, and matches the mean and covariance of these representations to offline reference statistics from high-quality speech. Unlike adversarial training, SR-FD requires no discriminator. Unlike post-generation re-ranking, SR-FD modifies the model directly and adds no inference-time computation.
The choice of representation matters, and a single statistic is unlikely to capture content drift well. We therefore formulate SR-FD over two frozen content representations: a Whisper semantic space and a connectionist temporal classification (CTC) posterior space. We use three complementary targets across these spaces. A Whisper target anchors successful low-step outputs. A CTC teacher target transfers the content behavior of a stronger ten-step sampler. A CTC real-speech target keeps the posterior distribution tied to natural speech. Together these targets penalize content drift under aggressive sampling compression.
The main empirical pattern is a reversal of the usual step-count tradeoff. Reducing the original VoxCPM2 from ten to four inference steps raises Seed-TTS English WER from 1.74\% to 2.23\%. Four-step SR-FD fine-tuning instead reaches 1.41\%, 36.5\% below the four-step baseline and 18.5\% below the ten-step baseline in relative terms; both gaps are significant under an utterance-level paired bootstrap, and the result is below the 1.47\% that ARCHI-TTS reports at 32 sampling steps~\cite{wu2026architts}.
Quality and speaker identity recover as well: UTMOS, DNSMOS, and Seed-TTS speaker similarity all return to near the ten-step level, and a blinded listening test finds no reliable listener preference against the ten-step baseline. An error analysis shows the WER gain comes mainly from fewer content substitutions across all prompt lengths. We therefore treat SR-FD primarily as an intelligibility-improving regularizer.
This paper makes three contributions:
\begin{itemize}\setlength{\itemsep}{1pt}
    \item \textbf{Distributional regularization for few-step TTS.} We formulate SR-FD as a Fr\'echet-distance training loss that matches the statistics of generated speech representations to offline reference statistics, using the same low-step sampler that will be used at inference time.
    \item \textbf{Multi-target content matching.} We combine three complementary targets across a Whisper semantic space and a CTC posterior space. This regularizes few-step content drift without adding inference-time computation.
    \item \textbf{Empirical and statistical analysis on VoxCPM2.} Four-step SR-FD fine-tuning reduces Seed-TTS English WER by 36.5\% relative to the four-step baseline and 18.5\% relative to the ten-step baseline, with significance confirmed by paired bootstrap, speaker similarity preserved, and a blinded listening test finding no reliable listener preference against the ten-step baseline. Error decomposition, length-bucket analysis, a target ablation, and an FD-diagnostic study localize where the improvement comes from.
\end{itemize}
\section{Related Work}
\textbf{Distributional metrics for generation.}
Fr\'echet Inception Distance (FID) compares image distributions by fitting Gaussians in a pretrained feature space \cite{heusel2017fid}, and Kernel Inception Distance replaces the Gaussian assumption with an unbiased maximum mean discrepancy (MMD) estimate \cite{binkowski2018kid}. Fr\'echet Audio Distance (FAD) ports the Gaussian comparison to audio with VGGish embeddings \cite{kilgour2018fad}; later work shows that the embedding choice strongly affects how well such distances track perception \cite{gui2024fadtk}, and kernel-based alternatives exist \cite{chung2024kad}. These distances are normally used for offline evaluation; SR-FD turns the same Gaussian comparison into a differentiable training loss and follows the embedding-sensitivity lesson by matching in two complementary content spaces.
\textbf{Distribution matching as a training signal.}
Training a generator to match feature statistics dates back to feature matching \cite{salimans2016gan} and MMD GANs \cite{li2017mmdgan}, which require a discriminator or an adversarially learned kernel. Score distillation \cite{poole2023dreamfusion,wang2023prolificdreamer} and distribution matching distillation \cite{yin2024dmd,yin2024dmd2} align a student generator with the distribution of a diffusion teacher and enable one-step image synthesis, but they need teacher score networks and sometimes an auxiliary fake-score model. SR-FD pursues the same goal with much lighter machinery: frozen speech encoders and precomputed first- and second-order moments, with no adversarial or learned component.
\textbf{Few-step diffusion and flow generation.}
Progressive distillation \cite{salimans2022progressive}, consistency models \cite{song2023consistency,song2023iCT}, rectified flow \cite{liu2023rectified}, and adversarial diffusion distillation \cite{sauer2023add} reduce sampling steps by retraining the model to imitate its own multi-step trajectory. Conditional flow matching \cite{lipman2023flowmatching} provides the stable base objective, but it still optimizes local velocity matching. These methods change how the sampler is trained; SR-FD is complementary, a distributional regularizer that directly targets what the few-step sampler produces.
\textbf{TTS generation paradigms.}
Codec language models such as VALL-E and SoundStorm generate discrete token sequences \cite{wang2023valle,borsos2023soundstorm}. Flow- and diffusion-based systems, including Voicebox, Matcha-TTS, NaturalSpeech 2 and 3, E2 TTS, and F5-TTS, generate continuous acoustic features \cite{le2024voicebox,mehta2024matchatts,shen2024naturalspeech2,ju2024naturalspeech3,eskimez2024e2tts,chen2024f5tts}, and VoiceFlow applies rectified flow for efficient TTS sampling \cite{guo2024voiceflow}. VoxCPM is tokenizer-free and autoregressive with a flow-matching acoustic decoder \cite{voxcpm2repo}; its continuous end-to-end audio path makes representation-level regularization natural. Preference-based post-training such as SpeechAlign \cite{zhang2024speechalign} also moves speech generation beyond pure likelihood; SR-FD differs in matching distribution statistics rather than learned preferences.
\textbf{Speech representations and adaptation.}
Self-supervised encoders such as wav2vec 2.0, HuBERT, and WavLM \cite{baevski2020wav2vec2,hsu2021hubert,chen2022wavlm}, and the weakly supervised Whisper model \cite{radford2023whisper}, are widely used as frozen judges of content and speaker identity. SR-FD uses a Whisper encoder and a wav2vec 2.0 CTC head as its frozen extractors; our evaluation uses WavLM-based speaker similarity \cite{chen2022wavlm} and the UTMOS and DNSMOS proxies \cite{saeki2022utmos,reddy2022dnsmos} under the Seed-TTS protocol \cite{anastassiou2024seedtts}. Low-rank adaptation (LoRA) \cite{hu2022lora} keeps the pretrained weights frozen during fine-tuning; SR-FD thus tests whether a distributional loss helps under parameter-efficient adaptation.
\section{Method}
\label{sec:method}
\begin{figure}[t]
\centering
\resizebox{\columnwidth}{!}{%
\begin{tikzpicture}[
  >={Stealth[round]},
  font=\scriptsize,
  node distance=4.5mm and 5mm,
  base/.style={draw, rounded corners=2pt, align=center, inner sep=2.5pt, minimum height=7mm},
  inp/.style={base, fill=black!4},
  trn/.style={base, fill=blue!12},
  frz/.style={base, fill=black!8},
  loss/.style={base, fill=orange!18}
]
\node[inp] (txt) {Text /\\prompt};
\node[trn, right=of txt] (samp) {Few-step sampler\\(VoxCPM2{+}LoRA)};
\node[inp, right=of samp] (gen) {Generated\\speech};
\node[frz, right=of gen] (ext) {Frozen extractors\\(Whisper, CTC)};
\node[base, right=of ext, fill=black!4] (fd) {Generated moments\\{+} Fr\'echet distance};
\node[loss, right=of fd] (srfd) {SR-FD\\loss};
\node[frz, below=3.5mm of fd] (ref) {Offline reference\\moments (3 targets)};
\draw[->] (txt) -- (samp);
\draw[->] (samp) -- (gen);
\draw[->] (gen) -- (ext);
\draw[->] (ext) -- (fd);
\draw[->] (fd) -- (srfd);
\draw[->] (ref) -- (fd);
\draw[->, dashed] (srfd.north) -- ++(0,3.2mm) -| node[pos=0.3, above=-0.5pt, font=\tiny] {gradient (training only)} (samp.north);
\end{tikzpicture}%
}
\caption{SR-FD training pipeline. The model samples speech with the deployment-time four-step sampler; frozen Whisper and CTC extractors map it to features whose mean and covariance are matched to offline reference moments via a Fr\'echet distance. Gradients flow only into the LoRA weights (dashed). Blue: trainable; gray: frozen; orange: loss.}
\label{fig:overview}
\end{figure}
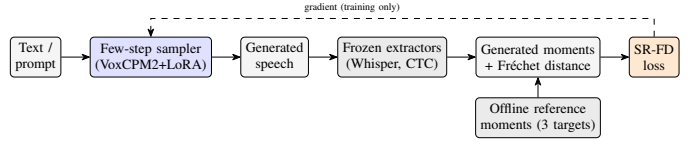
SR-FD augments the standard few-step TTS fine-tuning procedure with one
additional training objective. During fine-tuning, the model synthesizes speech with the same four-step
sampler used at deployment, and frozen speech encoders turn the generated audio
into feature vectors. We summarize each set of feature vectors by its mean
vector and covariance matrix,
and push the moments of generated speech toward reference moments computed
offline from desirable speech. The distance between the two moment sets is a
Fr\'echet distance, the same quantity behind FID and FAD, used here as a
differentiable loss. The loss needs no discriminator, adds no parameters, and is
removed at test time, so inference is unchanged. Figure~\ref{fig:overview} shows the pipeline.
\subsection{Base Model and Training Objective}
\label{sec:method_base}
VoxCPM2 is a tokenizer-free flow-matching autoregressive TTS model. Given an
input $x$, which contains the target text and an optional prompt audio with its
transcript, the model produces continuous AudioVAE latents and decodes them into
a waveform. The acoustic decoder is a diffusion transformer (DiT) trained with
conditional flow matching \cite{lipman2023flowmatching}.
Conditional flow matching works as follows. Take a latent $x_1$ extracted from
real speech and a noise sample $z\sim\mathcal{N}(0,I)$. Draw a time
$t\in[0,1]$ and form the point on the straight line between data and noise,
\begin{equation}
y = (1-t)\,x_1 + t\,z,
\qquad
v = z - x_1 ,
\end{equation}
where $v$ is the constant velocity along this line. The decoder predicts a
velocity $u_\theta(y,t)$, conditioned on the text and prompt, and the
flow-matching loss is the squared error
\begin{equation}
\mathcal{L}_{\mathrm{fm}}
=
\mathbb{E}_{x_1,z,t}
\left\|
u_\theta(y,t) - v
\right\|_2^2 .
\end{equation}
At inference, the decoder starts from noise and integrates the learned velocity
field with a small number of Euler steps; with only four steps this integration
is very coarse, exactly the regime our method targets.
The full fine-tuning objective combines the flow-matching loss with two other
parts:
\begin{equation}
\mathcal{L}_{\mathrm{base}}
=
w_{\mathrm{fm}}\,\mathcal{L}_{\mathrm{fm}}
+
w_{\mathrm{stop}}\,\mathcal{L}_{\mathrm{stop}}
+
\mathcal{L}_{\mathrm{aux}} .
\end{equation}
Here $\mathcal{L}_{\mathrm{stop}}$ is a stop-prediction loss: wrong stop
decisions cause truncation or runaway speech. $\mathcal{L}_{\mathrm{aux}}$
collects three small auxiliary losses inherited from the underlying recipe
(teacher-endpoint, preference-feature, and Whisper-text); their weights are
small, fixed in all runs (Section~\ref{sec:exp_opt}), and not a contribution of
this paper.
SR-FD adds one term on top:
\begin{equation}
\mathcal{L}
=
\mathcal{L}_{\mathrm{base}}
+
\lambda_{\mathrm{srfd}}\,\mathcal{L}_{\mathrm{srfd}} .
\end{equation}
The rest of this section defines $\mathcal{L}_{\mathrm{srfd}}$.
\subsection{Matching the Sampled-Speech Distribution}
\label{sec:method_match}
Standard fine-tuning supervises teacher-forced frames: the model sees the
ground-truth history and predicts the next frame, so the few-step sampler never
appears in the loss, and training can look healthy while the four-step sampler
drifts. SR-FD closes this gap by operating directly on sampled speech.
During each update, the model synthesizes a complete short utterance with the
deployment-time four-step sampler, keeping the computation differentiable. Let
$g_\theta(x_b)$ denote the generated waveform, that is, the audio that the
current model produces for input $x_b$ of mini-batch element $b$. Each frozen
extractor $\phi_k$ maps this generated audio to one utterance-level feature
vector,
\begin{equation}
\mathbf{h}_b^k
=
\phi_k\!\left(g_\theta(x_b)\right)
\in
\mathbb{R}^{d_k},
\end{equation}
where $d_k$ is the feature dimension of extractor $k$. The extractors are frozen and used only during training; SR-FD thus constrains
the distribution the sampler will actually produce at deployment, not the
supervised trajectory.
\subsection{Speech Representation Targets}
\label{sec:method_targets}
The final SR-FD configuration uses two frozen extractors and three reference
targets. The two extractors are a Whisper encoder \cite{radford2023whisper} for
semantic content and a wav2vec 2.0 CTC model \cite{baevski2020wav2vec2} for
phonetic content. All three targets describe speech content, because content
drift dominates the four-step failures in our experiments: the audio remains
speech-like, but an automatic speech recognition (ASR) system no longer recovers the intended words.
The first target is the \emph{low-step Whisper anchor}. We synthesize
utterances with the original four-step VoxCPM2, transcribe each generation with
an ASR system, keep only generations whose transcripts match the target text,
and pass them through a frozen Whisper large-v3 encoder, pooling the output into
a fixed utterance-level vector. The resulting statistics describe
what successful four-step outputs look like in a semantic ASR space, giving the
model a deployment-matched content anchor.
The second target is the \emph{teacher CTC target}, built from ten-step
VoxCPM2 teacher generations by summarizing wav2vec 2.0 CTC posteriors with mean and
standard-deviation pooling over non-blank content and blank statistics. The
ten-step teacher has lower WER, so this target transfers the content behavior of
a stronger sampler.
The third target is the \emph{real-speech CTC target}, which uses the same CTC
extractor with statistics from real LibriTTS voice-cloning speech, tying the CTC
branch to natural speech so the model does not simply copy teacher-specific
artifacts.
Table~\ref{tab:srfd_targets} summarizes the mixture;
Section~\ref{ssec:ablation} ablates each target.
\begin{table}[t]
\centering
\scriptsize
\caption{The three-target SR-FD reference mixture, computed offline and used
only through stored first- and second-order moments.}
\label{tab:srfd_targets}
\resizebox{\columnwidth}{!}{%
\begin{tabular}{llll}
\toprule
Target & Source & Extractor & Role \\
\midrule
Low-step Whisper anchor
& ASR-verified 4-step gen.
& Whisper
& Low-step content anchor \\
Teacher CTC target
& 10-step teacher gen.
& CTC
& Higher-step content transfer \\
Real-speech CTC target
& Real LibriTTS speech
& CTC
& Natural-speech grounding \\
\bottomrule
\end{tabular}%
}
\end{table}
\subsection{Reference and Generated Moments}
\label{sec:method_moments}
\textbf{Reference moments.}
For each target $j$ and extractor $k$, we precompute the reference moments from
an offline corpus. Let $\mathbf{h}_{i,j}^k=\phi_k(a_{i,j})$ denote the feature of
reference audio $a_{i,j}$ under extractor $k$. The stored statistics are
\begin{align}
\boldsymbol{\mu}_{r,j}^k
&=
\frac{1}{N_j}
\sum_{i=1}^{N_j}
\mathbf{h}_{i,j}^k, \\
\boldsymbol{\Sigma}_{r,j}^k
&=
\frac{1}{N_j}
\sum_{i=1}^{N_j}
(\mathbf{h}_{i,j}^k-\boldsymbol{\mu}_{r,j}^k)
(\mathbf{h}_{i,j}^k-\boldsymbol{\mu}_{r,j}^k)^\top .
\end{align}
Only these moments are stored; the reference audio is never used again. The two
feature spaces differ sharply in size: the CTC features are low-dimensional with
a well-conditioned covariance, while the Whisper features are high-dimensional
relative to the available sample count, so their covariance is rank-deficient. We therefore add a small diagonal term
$\epsilon I$ to the Whisper covariance and symmetrize the matrices before taking
the matrix square root. Because the absolute Fr\'echet value of the Whisper
branch is biased by this rank deficiency, we never select models by it; we
validate each target through the WER ablation in Section~\ref{ssec:ablation}.
Sample counts and feature dimensions are given in Section~\ref{sec:exp_data}.
\textbf{Generated moments with a feature queue.}
The Fr\'echet distance also requires the mean and covariance of generated features. A covariance estimated from a few utterances is statistically
unreliable, while generating hundreds of utterances per update is
computationally prohibitive. We therefore keep, for each extractor $k$, a queue
$\mathcal{Q}_t^k$ of features from recent updates. At step $t$, the generated
moments are computed over the queue together with the current mini-batch:
\begin{align}
\boldsymbol{\mu}_{g,t}^k
&=
\frac{1}{|\mathcal{Q}_t^k|}
\sum_{\mathbf{h}\in \mathcal{Q}_t^k}
\mathbf{h}, \\
\boldsymbol{\Sigma}_{g,t}^k
&=
\frac{1}{|\mathcal{Q}_t^k|}
\sum_{\mathbf{h}\in \mathcal{Q}_t^k}
(\mathbf{h}-\boldsymbol{\mu}_{g,t}^k)
(\mathbf{h}-\boldsymbol{\mu}_{g,t}^k)^\top
+
\epsilon I .
\end{align}
Features from earlier steps are detached; only the current mini-batch keeps
gradients. This design enables tractable moment matching: the queue
provides a population-scale covariance estimate at the memory cost of a
single batch, since the computation graph does not extend beyond the current step.
\subsection{The SR-FD Loss}
\label{sec:method_loss}
For each extractor $k$ and target $j$, SR-FD computes a Fr\'echet distance
between the generated and reference Gaussian moment estimates:
\begin{align}
\mathrm{FD}_{j,k}
&=
\|\boldsymbol{\mu}_{g,t}^k-\boldsymbol{\mu}_{r,j}^k\|_2^2
+
\operatorname{Tr}
(\boldsymbol{\Sigma}_{g,t}^k+\boldsymbol{\Sigma}_{r,j}^k)
\nonumber \\
&\quad
-
2\operatorname{Tr}
\left[
\left(
(\boldsymbol{\Sigma}_{r,j}^k)^{1/2}
\boldsymbol{\Sigma}_{g,t}^k
(\boldsymbol{\Sigma}_{r,j}^k)^{1/2}
\right)^{1/2}
\right].
\end{align}
Different feature spaces have different natural scales, so a raw sum over
targets would be dominated by whichever term is numerically largest. We
therefore divide each term by its own detached value:
\begin{equation}
\widetilde{\mathrm{FD}}_{j,k}
=
\frac{\mathrm{FD}_{j,k}}
{\stopgrad(\mathrm{FD}_{j,k})+\epsilon}.
\end{equation}
Each normalized term has magnitude near one, but its gradient still points in
the FD-reducing direction, so targets are balanced by gradient scale rather than
raw distance. The total loss
$\mathcal{L}_{\mathrm{srfd}}$ is a weighted average of the normalized terms,
first across targets within each extractor and then across extractors; with our
weights (Section~\ref{sec:exp_opt}), the Whisper and CTC branches contribute
equally and the two CTC targets split the CTC half.
A length gate admits a sample into the loss only when its duration is close to
the target duration, since strongly mismatched samples usually contain
truncation or runaway speech and matching moments on them injects noise;
per-utterance manifest weights further emphasize hard lexical examples
(thresholds in Section~\ref{sec:exp_opt}).
At test time the extractors, queues, reference moments, and Fr\'echet
computation are absent: the deployed model is a plain four-step VoxCPM2,
with no added parameters and no added inference computation.
\section{Experimental Setup}
\label{sec:setup}
\subsection{Data and Evaluation Protocol}
\label{sec:exp_data}
Fine-tuning uses a 767-row manifest derived from LibriTTS voice-cloning material. Each row contains the target text, continuation prompt information, and per-utterance training weights that emphasize hard lexical examples and protect teacher-consistent rows.
SR-FD reference statistics are computed offline from three sources: real LibriTTS voice-cloning speech, ten-step VoxCPM2 teacher generations, and ASR-verified four-step generations. The two CTC targets use 4,999 utterances each, at feature dimension $d=72$. The Whisper target uses 1,000 utterances, at feature dimension $d=960$; its covariance has a numerical rank of about 240, and we regularize it with $\epsilon=10^{-6}$ before taking the square root. Only the stored moments are loaded during training, and neither the manifest nor the reference statistics contain any Seed-TTS evaluation utterance.
Evaluation uses Seed-TTS English test-en \cite{anastassiou2024seedtts}. The full set contains 1,088 prompts, and the upstream Seed-TTS scorer reports 11,805 reference words. This upstream scorer is the primary WER metric: it is the external benchmark protocol. A 200-prompt gate subset with 2,070 reference words is used only for screening and for checkpoint selection in the target ablations.
\subsection{Systems Compared}
The backbone is VoxCPM2, a 2B-parameter tokenizer-free multilingual TTS model. We compare three systems under the same evaluation protocol: the pretrained model at four and at ten inference steps, and our four-step SR-FD fine-tuned model. We also include reported test-en results from two flow-matching systems, F5-TTS~\cite{chen2024f5tts} and ARCHI-TTS~\cite{wu2026architts}, quoting each system's best published configuration; their audio is unavailable under our pipeline, so we do not report similarity or proxies. ARCHI-TTS is the closest setting to ours, at its default 32 sampling steps. Large-scale post-trained language-model systems report 1.45\% (CosyVoice 3~\cite{du2025cosyvoice3}) and 1.24\% (Qwen3-TTS~\cite{hu2026qwen3tts}) under the same protocol; these rely on far larger corpora and post-training.
\subsection{Optimization and Decoding}
\label{sec:exp_opt}
Fine-tuning trains LoRA adapters with rank 32 and alpha 32 on the q, k, v, and o projections of the language-model and DiT components; the pretrained weights stay frozen. Training proceeds in two stages: a supervised LoRA adaptation trained with $\mathcal{L}_{\mathrm{base}}$ only, then 1,600 further steps with SR-FD enabled, using AdamW (weight decay 0.01, gradient-norm clip 0.03), bf16 precision, batch size 1, and a cosine learning rate from $3\times10^{-8}$ to zero with no warmup; checkpoints are saved every 100 steps. The second stage is a small residual refinement: the relative change in the LoRA weights is about $7.9\times10^{-5}$.
The loss weights are $w_{\mathrm{fm}}=0.006$ and $w_{\mathrm{stop}}=0.08$. The auxiliary weights are 0.001 for the teacher-endpoint loss, 0.001 for the preference-feature loss, and $10^{-5}$ for the Whisper-text loss; they are identical in every run. The SR-FD weight is $\lambda_{\mathrm{srfd}}=2\times10^{-4}$. The raw target weights are 1.0 for the Whisper anchor and 0.5 each for the teacher and real-speech CTC targets, with both extractor weights set to 1.0; the Whisper and CTC branches therefore contribute equally. The feature queue holds 50,000 vectors and includes the current mini-batch. The length gate accepts generated-to-target duration ratios in $[0.92, 1.08]$. During training, samples are generated with four-step Euler sampling, guidance 2.45, $\phi=0$, temperature 1.0, and sway 1.0, where $\phi$ and sway are VoxCPM2 sampler parameters; the training guidance matches the value used to construct the preference features and the sampled SR-FD target caches.
For evaluation, all four-step systems use true Euler decoding with guidance 2.35, $\phi=0$, temperature 1.0, sway 1.0, continuation prompting, maximum length ratio 6.0, and shared global seed 0; the ten-step baseline uses the same protocol except the number of inference steps. This protocol is frozen and shared across all systems we decode.
\subsection{Metrics and Selection}
The primary metric is upstream Seed-TTS English WER. We report both word-error counts and percentages because the strongest systems differ by only a few word errors. We assess paired differences with an utterance-level paired bootstrap over the 1,088 prompts, computed on per-utterance error counts derived from the upstream scorer outputs. For speaker identity, we report Seed-TTS SIM: the cosine similarity between WavLM-large speaker-verification embeddings \cite{chen2022wavlm} of the generated speech and the reference speaker audio, following the Seed-TTS protocol \cite{anastassiou2024seedtts}. For objective quality, we report UTMOS22-strong \cite{saeki2022utmos} and speechmos DNSMOS \cite{reddy2022dnsmos}, summarized by OVRL and P808 in the main tables. SIM, UTMOS, and DNSMOS are objective proxies; a blinded pairwise listening test provides the perceptual check (Section~\ref{ssec:listening}). Each trial pairs two outputs for the same Seed-TTS test-en item (same target text and prompt waveform): one from the four-step SR-FD model and one from the ten-step baseline. The A/B side assignment is randomized per listener and item, and thirteen listeners judge these pairs with explicit no-preference, both-bad, and skip options. The only variable within a pair is the generating system. We pre-specify a $\pm$10 percentage-point equivalence margin and assess it with two one-sided tests (TOST) \cite{schuirmann1987tost}.
The reported SR-FD model is the step-1600 checkpoint of the three-target run, selected on the full-set WER frontier; each target ablation instead selects its checkpoint on the 200-prompt gate subset and reports full-set numbers.
\section{Results and Analysis}
\label{sec:results}
\subsection{Main Seed-TTS Result}
Table~\ref{tab:main_results} gives the central comparison. Compressing the original VoxCPM2 from ten inference steps to four raises WER from 205/11805 = 1.74\% to 263/11805 = 2.23\%: the original four-step sampler is not simply a faster version of the ten-step system.
The four-step SR-FD model reverses this degradation. It reaches 167/11805 = 1.41\% WER, 36.5\% below the four-step baseline and 18.5\% below the ten-step baseline in relative terms, and below both reported flow-matching references in Table~\ref{tab:main_results}. Under the paired bootstrap, the reduction against the original four-step baseline is 0.81 percentage points with a 95\% confidence interval of $[0.57, 1.06]$ ($p<10^{-4}$), and the reduction against the ten-step baseline is 0.32 points with an interval of $[0.14, 0.51]$ ($p=0.0004$). Both claims are statistically supported; Sections~\ref{ssec:where} and~\ref{ssec:ablation} analyze where the improvement comes from.
\begin{table}[t]
\centering
\scriptsize
\caption{Main Seed-TTS English results (upstream scorer). SIM is Seed-TTS WavLM speaker similarity; UTMOS and DNSMOS are objective proxies, not human MOS. SR-FD is the three-target model at step 1600. Reported rows are flow-matching systems' best published test-en numbers; their audio is unavailable for SIM and proxy scoring.}
\label{tab:main_results}
\resizebox{\columnwidth}{!}{%
\begin{tabular}{lcccc}
\toprule
System & Steps & Upstream WER $\downarrow$ & SIM $\uparrow$ & UTMOS / DNSMOS OVRL / P808 $\uparrow$ \\
\midrule
VoxCPM2 & 4 & 2.23\% & 0.74 & 3.30 / 2.90 / 3.53 \\
VoxCPM2 & 10 & 1.74\% & 0.76 & 3.81 / 3.09 / 3.67 \\
VoxCPM2 + SR-FD (ours) & 4 & \textbf{1.41\%} & \textbf{0.76} & 3.76 / 3.07 / 3.65 \\
\midrule
F5-TTS~\cite{chen2024f5tts} (reported) & 32 & 1.83\% & -- & -- \\
ARCHI-TTS~\cite{wu2026architts} (reported) & 32 & 1.47\% & -- & -- \\
\bottomrule
\end{tabular}%
}
\end{table}
\subsection{Speaker Similarity, Quality Proxies, and Cost}
The WER gain costs neither speaker identity nor objective quality. SIM rises from 0.74 to 0.76, matching the ten-step baseline. UTMOS and DNSMOS recover most of the four-to-ten-step gap (Table~\ref{tab:main_results}). These proxies are not human ratings; Section~\ref{ssec:listening} tests the comparison perceptually.
Inference cost is unchanged by the loss: the deployed model is a four-step VoxCPM2 with LoRA adapters, with an aggregate real-time factor of 0.2285, about 8.6\% below the ten-step baseline at 0.2501. The wall-clock gain of step reduction is modest because the autoregressive component does not scale with flow steps, so we claim zero added inference cost rather than a large speedup.
\subsection{Blinded Listening Test}
\label{ssec:listening}
\begin{table}[t]
\centering
\caption{Blinded pairwise listening test, four-step SR-FD versus the ten-step baseline. Listeners show no reliable preference, and TOST supports practical equivalence within the pre-specified $\pm$10-point margin.}
\label{tab:listening}
\resizebox{\columnwidth}{!}{%
\begin{tabular}{lrl}
\toprule
Quantity & Value & Interpretation \\
\midrule
\multicolumn{3}{l}{\textit{Preference (13 listeners; 229 judgments; 128 decisive)}} \\
Wins, SR-FD : ten-step & 61 : 67 & near-even split \\
SR-FD preference & 47.7\% & near chance (50\%) \\
Exact binomial $p$ vs.\ 50\% & 0.659 & consistent with chance \\
Ties & 98/229 (42.8\%) & often no clear winner \\
Other non-decisive & 3/229 (1.3\%) & 2 both-bad, 1 skip \\
\midrule
\multicolumn{3}{l}{\textit{Equivalence test, pre-specified $\pm$10-point margin}} \\
Wilson 90\% CI (SR-FD pref.) & 40.5\%--54.9\% & inside the band \\
TOST at $\alpha=0.05$ & pass & equivalence supported \\
\midrule
\multicolumn{3}{l}{\textit{Robustness}} \\
Listener-clustered 90\% CI & 40.5\%--53.5\% & inside the band \\
High/normal-confidence votes & 58/120 (48.3\%) & still near even \\
\bottomrule
\end{tabular}%
}
\end{table}
The listening test asks whether listeners reliably separate the four-step SR-FD model from the ten-step baseline. They do not (Table~\ref{tab:listening}). The decisive-vote split is consistent with chance preference, and the Wilson 90\% confidence interval lies inside the pre-specified $\pm$10 percentage-point equivalence region, so TOST rejects non-equivalence at $\alpha=0.05$ \cite{schuirmann1987tost}. The claim is system-level practical equivalence under the $\pm$10-point margin, nothing tighter. The extra six inference steps of the baseline buy no reliable perceptual advantage.
\subsection{Where the Improvement Comes From}
\label{ssec:where}
\textbf{Error decomposition.}
Table~\ref{tab:sid} decomposes the errors by type. SR-FD reduces every error type relative to both baselines: substitutions fall from 203 (four-step) and 168 (ten-step) to 142, deletions from 43 to 21, and insertions stay rare. The model does not trade one error type for another.
\begin{table}[t]
\centering
\scriptsize
\caption{Error-type decomposition (per-utterance re-alignment of upstream
transcripts; totals can differ from the upstream aggregate by a few words).}
\label{tab:sid}
\begin{tabular}{lcccc}
\toprule
System & Steps & Sub. & Del. & Ins. \\
\midrule
VoxCPM2 & 4 & 203 & 43 & 18 \\
VoxCPM2 & 10 & 168 & 30 & 8 \\
VoxCPM2 + SR-FD (ours) & 4 & \textbf{142} & \textbf{21} & \textbf{5} \\
\bottomrule
\end{tabular}
\end{table}
\textbf{Prompt-length breakdown.}
Table~\ref{tab:buckets} splits the evaluation set by reference length. The SR-FD model is below both baselines in every bucket, with the largest gain over the four-step baseline on short prompts, where the four-step sampler also degrades most relative to ten steps.
\begin{table}[t]
\centering
\scriptsize
\caption{WER by reference length (words). Buckets partition the 1,088 prompts.}
\label{tab:buckets}
\resizebox{\columnwidth}{!}{%
\begin{tabular}{lccccc}
\toprule
Bucket & Words & N & Base 4 & Base 10 & SR-FD \\
\midrule
short  & $\le 10$ & 469 & 2.42\% & 1.73\% & \textbf{1.17\%} \\
medium & 11--12   & 303 & 2.62\% & 2.07\% & \textbf{2.01\%} \\
long   & $>12$    & 316 & 1.77\% & 1.50\% & \textbf{1.18\%} \\
\bottomrule
\end{tabular}%
}
\end{table}
\textbf{Qualitative cases.}
At the prompt level, the original four-step baseline renders ``stone towers'' as ``tone of voice'' and ``up to'' as ``approve''; SR-FD recovers both. SR-FD also introduces occasional regressions on prompts that both original baselines transcribe correctly, for example ``warty newt'' rendered as ``wordy newt'', ``its'' as ``his'', and one inserted trailing word. SR-FD shifts the error distribution toward fewer content substitutions, but it does not monotonically improve text fidelity on every prompt.
\subsection{FD Target Ablation}
\label{ssec:ablation}
Table~\ref{tab:target_ablation_compact3} reports a leave-one-out ablation of the
three reference targets. The all-target reference is the main model of
Table~\ref{tab:main_results}, the step-1600 three-target run at 167/11805
upstream errors. Each ablated model removes exactly one target, keeps the
remaining two active, and selects its best checkpoint on the same 200-prompt
gate subset. A removed target counts as useful if its best gate-selected checkpoint
remains worse on the full upstream evaluation.
All three targets contribute: removing the low-step Whisper anchor causes the
largest degradation, from 167 to 182 upstream word errors, so the
successful-output target is the main anchor for four-step intelligibility.
Removing the real-speech CTC target raises the errors to 176 and removing the
teacher CTC target to 175; the one-error difference is too small to rank them,
and keeping both gives the best full-set WER. Across all
ablations, UTMOS and DNSMOS stay within 0.005 of the full model, and removing the Whisper anchor even raises the
proxies slightly while WER worsens. The targets therefore act on
ASR-recoverable content, not on proxy quality scores.
Overall, the compact mixture is not redundant: all leave-one-out variants reach
the same gate error count of 23/2070 after checkpoint selection, yet all three
are worse than the all-target model on the full upstream test set.
\begin{table}[t]
\centering
\scriptsize
\caption{Leave-one-out ablation. Each run removes one target and selects its
checkpoint on the 200-prompt gate subset; UTMOS and DNSMOS stay within 0.005 of
the full model.}
\label{tab:target_ablation_compact3}
\resizebox{\columnwidth}{!}{%
\begin{tabular}{lccc}
\toprule
Removed target & Sel.\ step & Gate WER $\downarrow$ & Upstream WER $\downarrow$ \\
\midrule
None (all 3 targets)    & 1600 & 22/2070 & \textbf{1.41\%} \\
Low-step Whisper anchor & 1600 & 23/2070 & 1.54\% \\
Real-speech CTC         & 1000 & 23/2070 & 1.49\% \\
Teacher CTC             & 900  & 23/2070 & 1.48\% \\
\bottomrule
\end{tabular}%
}
\end{table}
\subsection{Is the Fr\'echet Distance Itself a Good Diagnostic?}
\label{ssec:fddiag}
SR-FD trains the model to reduce representation FD, so can the raw FD value
replace WER as a selection signal? The answer is no. Across the 16 saved checkpoints of the three-target run, the correlation
between raw FD scalars and gate WER is weak: the strongest association is the
real-speech CTC FD, with Spearman $\rho=0.383$ ($p=0.143$), and the Whisper
anchor FD shows no association at all ($\rho=-0.021$).
An independent CTC content-feature pass over 200 evaluation prompts shows the
same pattern at the system level. The empirical FD to real target speech
improves from 0.00795 for the original four-step baseline to 0.00681 for the
SR-FD model, so training does move generated features toward the reference; yet
the ten-step baseline is closer still at 0.00517 despite its higher WER. This
pass uses a separate extractor configuration from the training-time statistics,
so we treat it as an external sanity check. The conclusion is conservative: reducing representation FD during
training improves few-step intelligibility, but a smaller raw FD does not imply
a lower WER, and FD alone cannot select checkpoints or models.
\section{Conclusion}
This paper presented SR-FD, a training-time Speech Representation Fr\'echet Distance loss for tokenizer-free few-step TTS: the deployment-time four-step sampler generates speech whose frozen Whisper and CTC moments are matched to offline reference statistics from three content targets, with no discriminator and no inference cost.
On Seed-TTS English, four-step SR-FD fine-tuning cuts WER from the four-step baseline's 2.23\% to 1.41\%, below the ten-step baseline at 1.74\% and the 32-step 1.47\% ARCHI-TTS reference, with both gains bootstrap-significant, speaker similarity preserved, and a blinded listening test finding no reliable listener preference against the ten-step baseline. The gain consists mainly of fewer content substitutions, holds across all prompt lengths, and weakens whenever any reference target is removed; raw FD, in contrast, is a weak checkpoint selector, so external WER remains necessary. The broader lesson is to train the sampled-speech distribution, not only the teacher-forced trajectory, and to verify it with external metrics.
\balance
\bibliographystyle{IEEEtran}
\bibliography{merged}

\begin{thebibliography}{10}
\providecommand{\url}[1]{#1}
\csname url@samestyle\endcsname
\providecommand{\newblock}{\relax}
\providecommand{\bibinfo}[2]{#2}
\providecommand{\BIBentrySTDinterwordspacing}{\spaceskip=0pt\relax}
\providecommand{\BIBentryALTinterwordstretchfactor}{4}
\providecommand{\BIBentryALTinterwordspacing}{\spaceskip=\fontdimen2\font plus
\BIBentryALTinterwordstretchfactor\fontdimen3\font minus
  \fontdimen4\font\relax}
\providecommand{\BIBforeignlanguage}[2]{{%
\expandafter\ifx\csname l@#1\endcsname\relax
\typeout{** WARNING: IEEEtran.bst: No hyphenation pattern has been}%
\typeout{** loaded for the language `#1'. Using the pattern for}%
\typeout{** the default language instead.}%
\else
\language=\csname l@#1\endcsname
\fi
#2}}
\providecommand{\BIBdecl}{\relax}
\BIBdecl

\bibitem{le2024voicebox}
M.~Le, A.~Vyas, B.~Shi, B.~Karrer, L.~Sari, R.~Moritz, M.~Williamson,
  V.~Manohar, Y.~Adi, J.~Mahadeokar, and W.-N. Hsu, ``{Voicebox}: Text-guided
  multilingual universal speech generation at scale,'' in \emph{Advances in
  Neural Information Processing Systems (NeurIPS)}, 2023.

\bibitem{shen2024naturalspeech2}
K.~Shen, Z.~Ju, X.~Tan, Y.~Liu, Y.~Leng, L.~He, T.~Qin, S.~Zhao, and J.~Bian,
  ``{NaturalSpeech 2}: Latent diffusion models are natural and zero-shot speech
  and singing synthesizers,'' in \emph{International Conference on Learning
  Representations (ICLR)}, 2024.

\bibitem{ju2024naturalspeech3}
Z.~Ju, Y.~Wang, K.~Shen, X.~Tan, D.~Xin, D.~Yang, Y.~Liu, Y.~Leng, K.~Song,
  S.~Tang, Z.~Wu, T.~Qin, X.-Y. Li, W.~Ye, S.~Zhang, J.~Bian, L.~He, J.~Li, and
  S.~Zhao, ``{NaturalSpeech 3}: Zero-shot speech synthesis with factorized
  codec and diffusion models,'' in \emph{International Conference on Machine
  Learning (ICML)}, 2024.

\bibitem{chen2024f5tts}
Y.~Chen, Z.~Niu, Z.~Ma, K.~Deng, C.~Wang, J.~Zhao, K.~Yu, and X.~Chen,
  ``{F5-TTS}: A fairytaler that fakes fluent and faithful speech with flow
  matching,'' in \emph{Annual Meeting of the Association for Computational
  Linguistics (ACL)}, 2025.

\bibitem{voxcpm2repo}
\BIBentryALTinterwordspacing
Y.~Zhou, G.~Zeng, X.~Liu, X.~Li, R.~Yu, J.~Gui, J.~Wu, Z.~Wang, X.~Shen, R.~Ye,
  Z.~Zhang, J.~Zhou, B.~Bai, W.~Sun, M.~Deng, Q.~Shi, Z.~Wu, and Z.~Liu,
  ``Voxcpm2 technical report,'' 2026. [Online]. Available:
  \url{https://arxiv.org/abs/2606.06928}
\BIBentrySTDinterwordspacing

\bibitem{salimans2022progressive}
T.~Salimans and J.~Ho, ``Progressive distillation for fast sampling of
  diffusion models,'' in \emph{International Conference on Learning
  Representations (ICLR)}, 2022.

\bibitem{song2023consistency}
Y.~Song, P.~Dhariwal, M.~Chen, and I.~Sutskever, ``Consistency models,'' in
  \emph{International Conference on Machine Learning (ICML)}, 2023.

\bibitem{anastassiou2024seedtts}
P.~Anastassiou, J.~Chen, J.~Chen, Y.~Chen, Z.~Chen, Z.~Chen, J.~Cong, L.~Deng,
  C.~Ding, L.~Gao \emph{et~al.}, ``{Seed-TTS}: A family of high-quality
  versatile speech generation models,'' \emph{arXiv preprint arXiv:2406.02430},
  2024.

\bibitem{wu2026architts}
C.~Wu, J.~Deng, Z.~Liu, Z.~Dai, H.~He, and Q.~Kong, ``{ARCHI-TTS}: A
  flow-matching-based text-to-speech model with self-supervised semantic
  aligner and accelerated inference,'' \emph{arXiv preprint arXiv:2602.05207},
  2026.

\bibitem{heusel2017fid}
M.~Heusel, H.~Ramsauer, T.~Unterthiner, B.~Nessler, and S.~Hochreiter, ``{GANs}
  trained by a two time-scale update rule converge to a local {Nash}
  equilibrium,'' in \emph{Advances in Neural Information Processing Systems
  (NeurIPS)}, 2017.

\bibitem{binkowski2018kid}
M.~Bi{\'n}kowski, D.~J. Sutherland, M.~Arbel, and A.~Gretton, ``Demystifying
  {MMD GANs},'' in \emph{International Conference on Learning Representations
  (ICLR)}, 2018.

\bibitem{kilgour2018fad}
K.~Kilgour, M.~Zuluaga, D.~Roblek, and M.~Sharifi, ``{Fr{\'e}chet} audio
  distance: A reference-free metric for evaluating music enhancement
  algorithms,'' in \emph{Interspeech}, 2019.

\bibitem{gui2024fadtk}
A.~Gui, H.~Gamper, S.~Braun, and D.~Emmanouilidou, ``Adapting {Fr{\'e}chet}
  audio distance for generative music evaluation,'' in \emph{IEEE International
  Conference on Acoustics, Speech and Signal Processing (ICASSP)}, 2024.

\bibitem{chung2024kad}
Y.~Chung, P.~Eu, J.~Lee, K.~Choi, J.~Nam, and B.~S. Chon, ``{KAD}: No more
  {FAD}! an effective and efficient evaluation metric for audio generation,''
  \emph{arXiv preprint arXiv:2502.15602}, 2025.

\bibitem{salimans2016gan}
T.~Salimans, I.~Goodfellow, W.~Zaremba, V.~Cheung, A.~Radford, and X.~Chen,
  ``Improved techniques for training {GANs},'' in \emph{Advances in Neural
  Information Processing Systems (NeurIPS)}, 2016.

\bibitem{li2017mmdgan}
C.-L. Li, W.-C. Chang, Y.~Cheng, Y.~Yang, and B.~P{\'o}czos, ``{MMD GAN}:
  Towards deeper understanding of moment matching network,'' in \emph{Advances
  in Neural Information Processing Systems (NeurIPS)}, 2017.

\bibitem{poole2023dreamfusion}
B.~Poole, A.~Jain, J.~T. Barron, and B.~Mildenhall, ``{DreamFusion}:
  Text-to-{3D} using {2D} diffusion,'' in \emph{International Conference on
  Learning Representations (ICLR)}, 2023.

\bibitem{wang2023prolificdreamer}
Z.~Wang, C.~Lu, Y.~Wang, F.~Bao, C.~Li, H.~Su, and J.~Zhu, ``{ProlificDreamer}:
  High-fidelity and diverse text-to-{3D} generation with variational score
  distillation,'' in \emph{Advances in Neural Information Processing Systems
  (NeurIPS)}, 2023.

\bibitem{yin2024dmd}
T.~Yin, M.~Gharbi, R.~Zhang, E.~Shechtman, F.~Durand, W.~T. Freeman, and
  T.~Park, ``One-step diffusion with distribution matching distillation,'' in
  \emph{IEEE/CVF Conference on Computer Vision and Pattern Recognition (CVPR)},
  2024.

\bibitem{yin2024dmd2}
T.~Yin, M.~Gharbi, T.~Park, R.~Zhang, E.~Shechtman, F.~Durand, and W.~T.
  Freeman, ``Improved distribution matching distillation for fast image
  synthesis,'' in \emph{Advances in Neural Information Processing Systems
  (NeurIPS)}, 2024.

\bibitem{song2023iCT}
Y.~Song and P.~Dhariwal, ``Improved techniques for training consistency
  models,'' in \emph{International Conference on Learning Representations
  (ICLR)}, 2024.

\bibitem{liu2023rectified}
X.~Liu, C.~Gong, and Q.~Liu, ``Flow straight and fast: Learning to generate and
  transfer data with rectified flow,'' in \emph{International Conference on
  Learning Representations (ICLR)}, 2023.

\bibitem{sauer2023add}
A.~Sauer, D.~Lorenz, A.~Blattmann, and R.~Rombach, ``Adversarial diffusion
  distillation,'' in \emph{European Conference on Computer Vision (ECCV)},
  2024.

\bibitem{lipman2023flowmatching}
Y.~Lipman, R.~T.~Q. Chen, H.~Ben-Hamu, M.~Nickel, and M.~Le, ``Flow matching
  for generative modeling,'' in \emph{International Conference on Learning
  Representations (ICLR)}, 2023.

\bibitem{wang2023valle}
C.~Wang, S.~Chen, Y.~Wu, Z.~Zhang, L.~Zhou, S.~Liu, Z.~Chen, Y.~Liu, H.~Wang,
  J.~Li, L.~He, S.~Zhao, and F.~Wei, ``Neural codec language models are
  zero-shot text to speech synthesizers,'' \emph{arXiv preprint
  arXiv:2301.02111}, 2023.

\bibitem{borsos2023soundstorm}
Z.~Borsos, M.~Sharifi, D.~Vincent, E.~Kharitonov, N.~Zeghidour, and
  M.~Tagliasacchi, ``{SoundStorm}: Efficient parallel audio generation,''
  \emph{arXiv preprint arXiv:2305.09636}, 2023.

\bibitem{mehta2024matchatts}
S.~Mehta, R.~Tu, J.~Beskow, {\'E}.~Sz{\'e}kely, and G.~E. Henter,
  ``{Matcha-TTS}: A fast {TTS} architecture with conditional flow matching,''
  in \emph{IEEE International Conference on Acoustics, Speech and Signal
  Processing (ICASSP)}, 2024.

\bibitem{eskimez2024e2tts}
S.~E. Eskimez, X.~Wang, M.~Thakker, C.~Li, C.-H. Tsao, Z.~Xiao, H.~Yang,
  Z.~Zhu, M.~Tang, X.~Tan, Y.~Liu, S.~Zhao, and N.~Kanda, ``{E2 TTS}:
  Embarrassingly easy fully non-autoregressive zero-shot {TTS},'' in \emph{IEEE
  Spoken Language Technology Workshop (SLT)}, 2024.

\bibitem{guo2024voiceflow}
Y.~Guo, C.~Du, Z.~Ma, X.~Chen, and K.~Yu, ``{VoiceFlow}: Efficient
  text-to-speech with rectified flow matching,'' in \emph{IEEE International
  Conference on Acoustics, Speech and Signal Processing (ICASSP)}, 2024.

\bibitem{zhang2024speechalign}
D.~Zhang, Z.~Li, S.~Li, X.~Zhang, P.~Wang, Y.~Zhou, and X.~Qiu,
  ``{SpeechAlign}: Aligning speech generation to human preferences,'' in
  \emph{Advances in Neural Information Processing Systems (NeurIPS)}, 2024.

\bibitem{baevski2020wav2vec2}
A.~Baevski, H.~Zhou, A.~Mohamed, and M.~Auli, ``{wav2vec 2.0}: A framework for
  self-supervised learning of speech representations,'' in \emph{Advances in
  Neural Information Processing Systems (NeurIPS)}, 2020.

\bibitem{hsu2021hubert}
W.-N. Hsu, B.~Bolte, Y.-H.~H. Tsai, K.~Lakhotia, R.~Salakhutdinov, and
  A.~Mohamed, ``{HuBERT}: Self-supervised speech representation learning by
  masked prediction of hidden units,'' \emph{IEEE/ACM Transactions on Audio,
  Speech, and Language Processing}, vol.~29, pp. 3451--3460, 2021.

\bibitem{chen2022wavlm}
S.~Chen, C.~Wang, Z.~Chen, Y.~Wu, S.~Liu, Z.~Chen, J.~Li, N.~Kanda,
  T.~Yoshioka, X.~Xiao, J.~Wu, L.~Zhou, S.~Ren, Y.~Qian, Y.~Qian, J.~Wu,
  M.~Zeng, X.~Yu, and F.~Wei, ``{WavLM}: Large-scale self-supervised
  pre-training for full stack speech processing,'' \emph{IEEE Journal of
  Selected Topics in Signal Processing}, vol.~16, no.~6, pp. 1505--1518, 2022.

\bibitem{radford2023whisper}
A.~Radford, J.~W. Kim, T.~Xu, G.~Brockman, C.~McLeavey, and I.~Sutskever,
  ``Robust speech recognition via large-scale weak supervision,'' in
  \emph{International Conference on Machine Learning (ICML)}, 2023.

\bibitem{saeki2022utmos}
T.~Saeki, D.~Xin, W.~Nakata, T.~Koriyama, S.~Takamichi, and H.~Saruwatari,
  ``{UTMOS}: {UTokyo-SaruLab} system for {VoiceMOS} challenge 2022,'' in
  \emph{Interspeech}, 2022.

\bibitem{reddy2022dnsmos}
C.~K.~A. Reddy, V.~Gopal, and R.~Cutler, ``{DNSMOS P.835}: A non-intrusive
  perceptual objective speech quality metric to evaluate noise suppressors,''
  in \emph{IEEE International Conference on Acoustics, Speech and Signal
  Processing (ICASSP)}, 2022.

\bibitem{hu2022lora}
E.~J. Hu, Y.~Shen, P.~Wallis, Z.~Allen-Zhu, Y.~Li, S.~Wang, L.~Wang, and
  W.~Chen, ``{LoRA}: Low-rank adaptation of large language models,'' in
  \emph{International Conference on Learning Representations (ICLR)}, 2022.

\bibitem{du2025cosyvoice3}
Z.~Du, C.~Gao, Y.~Wang, F.~Yu, T.~Zhao, H.~Wang, X.~Lv, H.~Wang, C.~Ni, X.~Shi,
  K.~An, G.~Yang, Y.~Li, Y.~Chen, Z.~Gao, Q.~Chen, Y.~Gu, M.~Chen, Y.~Chen,
  S.~Zhang, W.~Wang, and J.~Ye, ``{CosyVoice 3}: Towards in-the-wild speech
  generation via scaling-up and post-training,'' \emph{arXiv preprint
  arXiv:2505.17589}, 2025.

\bibitem{hu2026qwen3tts}
H.~Hu, X.~Zhu, T.~He, D.~Guo, B.~Zhang, X.~Wang, Z.~Guo, Z.~Jiang, H.~Hao,
  Z.~Guo, X.~Zhang, P.~Zhang, B.~Yang, J.~Xu, J.~Zhou, and J.~Lin,
  ``{Qwen3-TTS} technical report,'' \emph{arXiv preprint arXiv:2601.15621},
  2026.

\bibitem{schuirmann1987tost}
D.~J. Schuirmann, ``A comparison of the two one-sided tests procedure and the
  power approach for assessing the equivalence of average bioavailability,''
  \emph{Journal of Pharmacokinetics and Biopharmaceutics}, vol.~15, no.~6, pp.
  657--680, 1987.

\end{thebibliography}
\end{document}